\documentclass[sigconf]{acmart}

\acmConference[PoliSim@CHI 2026]{PoliSim@CHI 2026: LLM Agent Simulation for Policy}{April 16, 2026}{Barcelona, Spain}
\renewcommand{\footnotetextcopyrightpermission}[1]{%
  \footnotetext{%
    This paper was prepared for \href{https://polisim.net/}{PoliSim@CHI 2026: LLM Agent Simulation for Policy}, a workshop at \href{https://chi2026.acm.org/}{CHI 2026 (CHI Conference on Human Factors in Computing Systems)}, April 16, 2026, Barcelona, Spain.
  }
}

\begin{document}

\title{We Need Strong Preconditions For Using Simulations In Policy}

\author{Steven Luo}
\affiliation{%
  \institution{University of California, Berkeley}
  \city{Berkeley}
  \state{California}
  \country{USA}
}
\email{sfluo@berkeley.edu}

\author{Saanvi Arora}
\affiliation{%
  \institution{University of California, Berkeley}
  \city{Berkeley}
  \state{California}
  \country{USA}
}
\email{saanvia@berkeley.edu}

\author{Carlos Guirado}
\affiliation{%
  \institution{University of California, Berkeley}
  \city{Berkeley}
  \state{California}
  \country{USA}
}
\email{guirado@berkeley.edu}

\begin{abstract}
Simulations, and more recently LLM agent simulations, have been adopted as useful tools for policymakers to explore interventions, rehearse potential scenarios, and forecast outcomes. While LLM simulations have enormous potential, two critical challenges remain understudied: the dual-use potential of accurate models of individual or population-level human behavior and the difficulty of validating simulation outputs. In light of these limitations, we must define boundaries for both simulation developers and decision-makers to ensure responsible development and ethical use. We propose and discuss three preconditions for societal-scale LLM agent simulations: 1) do not treat simulations of marginalized populations as neutral technical outputs, 2) do not simulate populations without their participation, and 3) do not simulate without accountability. We believe that these guardrails, combined with our call for simulation development and deployment reports, will help build trust among policymakers while promoting responsible development and use of societal-scale LLM agent simulations for the public benefit.
\end{abstract}



\maketitle

\section{Introduction}
Simulations have long been used to tackle large-scale questions in policy domains such as transportation, public health, and economics. With recent advances in the capabilities of large language model agent simulations \cite{hewitt2024predicting}\cite{binz2025foundation}\cite{10.1145/3586183.3606763}\cite{park2024generative}, many researchers have identified promising directions for applying LLM simulations to societal-scale problems \cite{anthis2025llm}, particularly where modeling language-mediated behavior and tacit knowledge is important.

We illustrate the promises — and the potential perils — of the same LLM simulation in two example scenarios:

\begin{enumerate}
    \item A simulation modeling attendees at a large event.
    \begin{enumerate}
        \item Promise: Improve emergency preparedness and develop better evacuation procedures by finding blind spots or possible failure points before any crisis ever occurs. \cite{li2025makes}
        \item Peril: Allow malicious actors to strategically maximize the disruption and harm they cause.
    \end{enumerate}
    \item A simulation modeling the behavioral dynamics of immigrant populations in a country.
    \begin{enumerate}
        \item Promise: Help test how different social support policies might affect outcomes like employment, education, housing stability, or access to healthcare to design more effective services for immigrant communities.
        \item Peril: Enable nativist leaders to better exploit how immigrant populations respond to different immigration enforcement operations to find a maximally exclusive strategy for deportations; use certain forecasts to justify harsher border controls or unequal treatment of certain national, ethnic, or religious groups.
    \end{enumerate}
\end{enumerate}

In both cases, the same technology is being used, but the difference lies in the power dynamics between the institutions using and interpreting simulated outputs and the populations being simulated. However, we are not aware of widely adopted or established boundaries governing when and how these simulations should be deployed.

We highlight dual-use and validity concerns because they identify the two most fundamental ways societal-scale LLM agent simulations can cause harm. A simulation may be well-developed and useful for policy, but can be dangerous in the hands of bad actors; conversely, a simulation may be developed with helpful intents, but can cause harm by misleading decision-makers if it is untested. In other words, good intentions or technical sophistication alone are insufficient to justify decision-maker adoption, so any governance framework for these tools must address both how simulations can be misused and how they can be wrong.

\subsection{Dual-Use Concerns}
The dual-use argument surrounding LLM agent simulations is not unique to this technology. In this context, dual-use refers to the potential for technology to generate substantial social benefit while also being used in harmful ways \cite{harris2016governance}. This dilemma has been discussed long before the first large language models were released, such as in nuclear and life science research \cite{nouri2012growing}. 

LLM agent simulations extend the dual-use potential of AI systems: they can help policymakers test interventions and anticipate second-order effects, but can also optimize mechanisms for manipulation and social control at a societal-level. More broadly, the dual-use literature on LLMs argues that capabilities developed for beneficial assistance can also reduce the cost and expertise required for harmful activity, making it easier for anyone to carry out scams while also attracting more sophisticated adversaries \cite{brundage2024malicioususeartificialintelligence} and suggesting that the policy value of these systems cannot be separated from their misuse potential.

\subsection{Validity Concerns}
Simulations can be powerful, but only when we can trust them to reflect reality; we struggle to ensure that simulated agents exhibit realistic behaviors rather than stereotypes \cite{wang2025large}\cite{ng2025llm}, scale simulations to capture emergent population-level dynamics \cite{10.5555/3709347.3743565}, and decide when to trust simulated outcomes to inform real decisions \cite{wang2025llmbasedhumansimulationsreliable}. 

Recent commercial platforms offering synthetic audiences highlight growing market interest \cite{vranica2026aaru}, but many of these products are optimized for short-horizon marketing and design questions with proprietary black-box agents, and have limited support for rigorous calibration as well as uncertainty quantification across domains and interoperability with existing data platforms and systems. As a result, these systems provide outputs that are closer to one-off predictions about audience response rather than transparent, scientifically grounded methods for aligning agent behavior with human data and for knowing when simulations are reliable enough to inform high-stakes policy decisions.

When using LLM agent simulations to forecast potential future events, validation is how we distinguish a principled decision making tool from a ``Magic 8 Ball''. However, validation in the form of prediction accuracy from classical machine learning does not work due to our inability to observe counterfactuals that never happen in the real world, and clever workarounds to selectively labeled data \cite{10.1145/3097983.3098066} do not work or are not applicable for these prediction tasks. There is clearly an urgent need to address the validation challenge present in generative social simulation \cite{larooij2025validation}, such as through network science approaches or by interrogating the available evidence from simulation scenarios.

The dual-use challenge means that we cannot assume good intent will govern how simulations are used. The validity challenge means we cannot rely on technical rigor to catch harmful outputs before they reach policymakers. Taken together, it is clear that simulations can cause serious harm through both misuse and honest error, with no reliable mechanism to prevent either. In the absence of such, the field needs explicit boundaries: not on the technology itself, but on the conditions under which simulations are permitted to inform consequential decisions. We propose three such preconditions.

\section{Three Preconditions}
Each of the following preconditions describe a failure that becomes possible when decision-makers accept simulation outputs as authoritative answers without questioning the processes that led to the outputs they see. We believe that these preconditions can mitigate the impacts of technocratic overreach, which at its worst may substitute simulations for the critical political and deliberative processes through which affected communities have standing to shape decisions about their lives. 

\subsection{Do Not Treat Simulations of Marginalized Populations as Neutral Technical Outputs}

The standard critique of biased models focuses on their accuracy, purporting that the model performs worse for marginalized groups — or groups positioned at the social periphery through institutionalized patterns of exclusion and inequality \cite{fluit2024social} — because they are underrepresented in training data. The deeper issue at hand is representational harm, not allocative harm. While training data reflects historical patterns of inequality, this is not a problem that better data alone can solve, since the harm is not that the model is inaccurate but rather that the model enters inequality into the policy record as a technical finding. Representational validity also remains a concern as LLMs tend to stereotype subpopulations \cite{cheng-etal-2023-marked}\cite{lutz2025promptmakespersonasystematic}, and the problem of machine learning systems causing harm to marginalized groups does not begin and end with training data and representation, from the “digital poorhouse” \cite{eubanks2018automating} to the “New Jim Code” \cite{benjamin2019race}.

This is made worse by the fact that discriminatory outcomes encoded in technical systems carry a legitimacy that explicit discrimination does not, because they appear to be the product of neutral methodology rather than political choice \cite{Green_2021}. At the same time, there is no independent ground truth against which to validate a simulation of marginalized population behavior, because the historical data that would serve as the “ground truth” is itself a product of discriminatory policy \cite{mayson2019bias}. This feedback loop makes this issue unresolvable through data collection alone. 

Simulations that claim to predict the behavior of marginalized communities, broken down by race, income, immigration status, or disability, or other dimensions of marginalization should not serve significant roles in policy decisions unless validation was conducted against community-reported experiences beyond historical data, calibration data was generated through a participatory process with the communities involved, and the model explicitly discloses the heterogeneity that it cannot capture, including a description of which behavioral and adaptive circumstances fall outside the model’s scope. Where these conditions cannot be met, the simulation should be scoped to populations for which validation is feasible, and its inapplicability to excluded groups must appear prominently in the policy record. 

Critical to this discussion is the recognition that even when a simulation is acknowledged as imperfect or provisional, once it enters a policy record it can acquire institutional authority and become a reference point for later decisions, as the imperfect output eventually is treated as an established baseline against which future models are calibrated. Therefore, policymakers must be extremely cautious about the scope and applicability of simulations to policy questions involving sensitive attributes, as the legitimization of a problematic simulation via a policy record creates a direct mechanism for compounding harms. 

\subsection{Do Not Simulate Populations Without Their Participation}

Participatory design scholarship distinguishes between two relationships a community can have with a research or policy process: either as a subject, whose behavior is observed, modeled, and acted upon, and as a participant, whose knowledge, priorities, and interpretive authority shape the process \cite{10.1145/3617694.3623228}. LLM agent simulations that treat affected communities as subjects, rather than participants, are extractivist precisely when they derive value from a community’s experience without return, consent, or meaningful participation in how that experience is used. This extractivism creates both ethical and epistemic issues as affected communities might identify model failures that technical reviewers miss, because they know from lived experience when a model’s assumptions do not match how people in their community actually behave or make decisions \cite{10.1145/3551624.3555290}. 

That said, not all participation is equal: there is a meaningful difference between asking a community to review a simulation’s outputs after it has been run (satisfying the form of participation) and involving that community in designing scenarios, choosing whose behavior gets modeled, and which outcomes are measured (satisfying both the form and substance of participation). Participation must be constitutive, where participants are given meaningful opportunities to shape the terms of the process, not consultative, where participants merely provide feedback \cite{smith2026reimagining}\cite{10.1145/3551624.3555285}\cite{delgado2021stakeholderparticipationaiadd}. 

There is a deeper concern here regarding the strength of our deliberative democracy. When simulations are introduced into policy spaces, they tend to restructure deliberative processes; ultimately all policy decisions, especially those related to resource allocation, are fundamentally political and involve competing values, contested tradeoffs, and questions about what a society owes its citizens. These are questions that traditional democratic deliberation is designed to surface and negotiate. The introduction of a simulation into that process might reframe those questions as technical ones, moving authority away from accountable policymakers and deliberative processes towards modeling teams and the parties that commission them \cite{10.1093/oxfordhb/9780190067397.013.15}. This results in an epistemological asymmetry between those who can interrogate the model’s assumptions and retains the power to contest its conclusions, and those who cannot and therefore must defer to them. Legislators, agency heads, and community members often lack the technical knowledge needed to fully understand and interrogate these kinds of simulations; therefore, the simulation does not inform their judgment but rather substitutes for it. 

Simulations must satisfy participation requirements around design, validation, and interpretation. Satisfying design participation requires co-design, such that affected communities have substantive input into the selection of scenarios, the measurement of outcomes, and behavioral assumptions before the model is built. Validation participation requires the model to be validated against community-reported experience, not only against historical administrative data, with the understanding that divergences between a model’s outputs and community’s reports is a model failure. Finally, participation requires that before outputs inform policy recommendations, affected communities have access to simulation results in plain language, a clear account of what the model cannot capture, and a formal opportunity to contest its interpretation. 

\subsection{Do Not Simulate Without Accountability}

It is important to recognize that when technology is used to shape any consequential policy decision, it creates a pathway for harm with no clear responsible party. Each actor involved (the model developer, the commissioning party, the policymaker) places blame on the other. Affected communities are therefore left with no legible account of how the decision was made or who made it. The use of simulations to inform policy enables the black box of “technology” to ultimately absorb the decision while distributing the responsibility across several actors, one of the features that makes simulations potentially politically appealing. Existing legal frameworks, including equal protection doctrine and disparate impact analysis in the U.S. legal regime, provide potential remedies when policy decisions produce outcomes that fall hardest on protected groups \cite{griggs}\cite{beyondintent2021}. However, by the time a simulation’s output enters a rulemaking or legislative process, the decisions (e.g. which scenarios to run, which outcomes to measure, which populations to disaggregate) are largely invisible. Those legal frameworks require a traceable decision attributable to an identifiable actor; tracing the allocation of a burden or benefit back to a simulation with undisclosed design choices leaves affected parties with no viable doctrinal hook \cite{38145d6a-c339-3557-aff4-54ebe765b0d9}. 

This precondition is not just about transparency in the disclosure sense, but about building accountability into the structure of how simulations are commissioned, validated, and acted upon \cite{ananny2018seeing}. First, the decision chain must be legible. Every consequential choice in the simulation process (e.g. who commissioned it, what scenarios were selected, who validated it, what the model cannot capture) must be attributable to an identified party and part of the official record. Second, the validation must be independent. The party that commissions a simulation should not control its validation, especially where the commissioning party has interests in particular outputs. Finally, recourse for affected parties should be clearly defined. Without a formal mechanism for contesting the features of the simulation and its use in the policy arena, the participation requirements as defined in the previous preconditions are unenforceable and the legal hooks in the equal protection and disparate impact doctrine remain inaccessible. 

\section{Discussion}
Setting these preconditions now is in the best interest of both decision-makers and developers. For decision-makers, our preconditions provide a framework for using simulations without undermining accountability or public trust. For developers, our preconditions offer a way to set clear boundaries around acceptable use and to avoid a race to the bottom driven by competitive or political pressure. Because the adoption of LLM agent simulations in policymaking is likely to be slow and research in this area is relatively nascent, there is an opportunity to establish norms early before harmful practices harden into standard practice and shape supply and demand in ways that support responsible, trustworthy use.

\subsection{Diffusion}
While the speed of diffusion of AI technologies has varied between applications \cite{challapally2025genai}, diffusion in safety-critical areas and the emergence of transformative economic and societal impacts has been relatively slow \cite{narayanan2025ainormal}. There are several reasons to believe that the diffusion of LLM agent simulations to inform policy will be significantly slower than other applications of general-purpose AI systems. Policymaking is a fundamentally trust and relationship-driven process; while using scientific findings to inform policy development is relatively standard practice \cite{bommasani2025advancing}\cite{bommasani2025california}, trust in technology as a tool for evidence generation is more mixed and attitudes on the promise of computational tools can vary based on partisan affiliation and age \cite{furnas2025partisan}. For example, the 2021 Facebook hearings in the U.S. Senate Commerce Subcommittee on Consumer Protection, Technology, and Data Privacy highlighted gaps in decision-makers’ understanding of how basic technologies function \cite{wise2021finsta}, though upskilling programs targeted at graduate students, scientists, and legislative staff at different levels of government are helping to improve policymaker fluency in AI \cite{brown2025}\cite{stanfordhai_congressional_boot_camp_ai}\cite{ccst_legislative_staff_academy_ai}.

While LLM agent simulations have the potential to improve decision quality, AI technologies remain broadly unpopular among the American public. In a recent poll, only 26\% of voters held positive views towards AI, and generally do not trust either of the major political parties to effectively handle this emerging technology \cite{smith2026poll}. Elected decision-makers are also particularly sensitive to the traceability of their decisions \cite{mildenberger2025effect}, and the negative political ramifications of using faulty simulations or appearing to delegate decision accountability to an uninterpretable computation model would likely lead to losses of support across all levels of their constituencies regardless of their “home style” \cite{fenno1977us}.

The convergence of technical hurdles as well as political and public skepticism create a unique set of circumstances that suggest a slower, more hesitant path to adopting simulation tools to inform policy, particularly at the highly visible federal levels of government. However, while this slow diffusion and adoption process limits the size of the developer and user base of LLM agent simulations, the smaller community around research and application makes even voluntary guidelines more effective in shaping responsible futures of this modeling tool. It is critical that developers of LLM agent simulations as well as decision-makers who adopt these tools to assist in policymaking keep the validity, dual-use, and accountability concerns front and center during all stages of research, development, and deployment.

\subsection{Building Trust}
Trust is an essential implementation factor \cite{10.1145/3531146.3533158} in effectively using societal-scale simulations to inform policy — the decision-maker needs to trust that the tool they are using will provide them with good information, and the developer needs to trust that the user will not misuse their tool in an “off-label”, potentially harmful way. Trust in applications in emerging technologies is also particularly fragile because of the political capital expended by early adopters in breaking with the status quo in their organization; if the pilot fails or causes harm, reputational damage can make continued use and wider adoption difficult or entirely impossible \cite{olsen2021minding}. 

In order for LLM agent simulations to be politically acceptable in the long-term, developers of LLM agent simulations must consider the preconditions while being especially careful of the validation mechanisms used. Deploying a faulty simulation, particularly in high-stakes environments, can transform a neutral computational tool into a negatively charged, politically salient issue. In fact, historical cases of emerging technologies suggest that early visible failures can have lasting effects on public trust and political legitimacy. While the stakes differ substantially, the Three Mile Island accident is an example of a failure that shows how a single failure can harden skepticism and slow future adoption beyond the context of the original mistake. Opposition to nuclear power steadily rose after the accident \cite{osti_5354975}, and it took more than forty years for public sentiment to recover \cite{brenan2025nuclear}. 

Developers of LLM agent simulations therefore must be especially careful to validate their claims about what their simulations are capable of and the societal questions they are actually able to help answer. Repeatedly overpromising and underdelivering will rapidly compromise efforts to build trust in the researcher-policymaker relationship; trust in science and the researchers that seek to advise policy efforts is crucial for long-term efforts to bridge the gap between researchers and decision-makers in shaping public policy \cite{chan2025enhancing}. For developers, understanding why or why not certain decision-makers adopt AI — such as tools like LLM agent simulations — and personalizing demonstrations as well as validated use cases to respond to their specific concerns can facilitate trusted adoption \cite{yu2025decisionmakersnotuse}, while decision-makers interested in using simulations can assess whether the tool is appropriate for their use by interrogating development processes and claims about model results \cite{smits2023using}.

\subsection{Simulation Development and Deployment Reports}
Using LLM agent simulations as a source of information in the policy development process inherently outsources accountability from human experts and human-published research to evidence generated by a computational model with accountability spread across developers, commissioners, and decision-makers \cite{kemper2019transparent}\cite{besio2025algorithmic}. Consistent with our third precondition, clearly identifying who holds accountability for how simulations are commissioned, validated, and acted upon is essential to fostering public trust for using simulations to inform policy decisions. 

Drawing on model cards detailing performance characteristics about the released machine learning model \cite{10.1145/3287560.3287596}, datasheets for datasets \cite{10.1145/3458723}, and dataset nutrition labels \cite{holland2020dataset}, we call on LLM agent simulation developers to release development and deployment reports that describe why certain design choices were made as well as narratives around the adoption and downstream use of the simulation. These “simulation development and deployment reports” should answer questions including: \textit{Who built the simulation? How was it built and who was consulted? How was the simulation validated, and why were these validation measures sufficient to prove that the simulation is an accurate model of the population it tries to represent? Who should be interpreting the outputs? How should the outputs be interpreted and used? How has the simulation been used, and what were some of the successes and failures in deployment?} 

Unlike general-purpose AI models, LLM agent simulations may be built to model particular domains, behaviors, or attributes in society, requiring particular skillsets to interpret outputs. Carefully specifying who should be interpreting simulation outputs and how they can be used will be important to preventing harms stemming from misinterpretations of outputs, while intentionally defining which humans must be kept in the loop. A singular static report release is also not enough, as simulation development and deployment reports should also be continually updated by developers as they collaborate with new decision-makers and apply their simulation to new problems to detail the origins of collaborations as well as newly surfaced considerations and lessons learned. Not only will these reports facilitate thorough interrogation of potential use cases to prevent misuse by decision-makers and of development and validation processes to prevent the adoption of faulty simulations, but documenting the history of deployments will help members of the public remain active participants in the processes of deliberative democracy.

\section{Conclusion}
Societal-scale LLM agent simulations have the potential to become useful tools for exploring counterfactual outcomes and informing policymaking processes, but their legitimacy cannot rest on technical novelty alone. Because these systems can be misused and fail to accurately represent the populations they try to model, they should not inform consequential decisions without clear boundaries on how they are built, validated, and deployed. We argue for three such preconditions: do not treat simulations of marginalized populations as neutral technical outputs, do not simulate populations without their participation, and do not simulate without accountability. Together, these guardrails and our call for simulation development and deployment reports offer a path forward to ensure that LLM agent simulations for policy are used in ways that are trustworthy and accountable to developers, decision-makers, and the broader public.

\begin{acks}
We thank Ro Encarnación, Princess Sampson, Lauren Chambers, and Serena Booth for their insightful comments and thoughtful suggestions during the writing process.
\end{acks}

\bibliographystyle{ACM-Reference-Format}
\bibliography{polisim_workshop}

\end{document}